\documentclass[12pt]{article}
\usepackage[utf8]{inputenc}
\usepackage{amsmath, amsthm, amssymb, amsfonts}
\usepackage{import}
\usepackage[authoryear]{natbib}

\usepackage{fullpage}
\usepackage{authblk}
\usepackage[bibliography=common]{apxproof}
\usepackage{import}
\usepackage{mathtools}
\usepackage{tikz}
\usepackage{float}
\usepackage{enumitem}
\usepackage{bbm}
\usepackage[colorlinks=false,pagebackref=true, hidelinks]{hyperref}
\usetikzlibrary{calc,patterns,angles,shapes, positioning, intersections, quotes}

\newtheoremrep{thm}{Theorem}
\newtheoremrep{lem}{Lemma}
\newtheoremrep{prop}{Proposition}

\newtheoremrep{proposition}{Proposition}

\usepackage{graphicx}
\newtheorem{corollary}{Corollary}
\newtheorem{fact}{Fact}

\newtheorem{example}{Example}

\newtheorem{remark}{Remark}
\theoremstyle{definition}
\newtheorem{definition}{Definition}

\renewcommand{\P}{\mathcal{P}}
\newcommand{\D}{\mathcal{D}}
\newcommand{\X}{\mathcal{X}}
\newcommand{\rR}{\mathrel{R}}
\newcommand{\rP}{\mathrel{P}}
\newcommand{\f}{\varphi}
\newcommand{\Diff}{\mathrm{Diff}}

\newcommand{\df}[1]{\textit{#1}}

\allowdisplaybreaks

\date{\today\\\small{\href{https://goelsumit.com/files/exchange_ttc.pdf}{(Link to latest version)}}}

\begin{document}

\title{TTC Domains\thanks{We are grateful to Julien Combe, Federico Echenique, Arunava Sen, William Thomson, Huaxia Zeng, as well as participants at the Conference on Economic Design at University of Essex (2025), WINE 2025, Annual Conference at Indian Statistical Institute Delhi (2025), Durham Economic Theory Conference (2026), Conference on Mechanism and Institution Design at University of York (2026), and seminar audiences at NYU Abu Dhabi for helpful comments and suggestions. Tamura acknowledges financial support from LabEx Ecodec (ANR-11-LABX-0047). An extended abstract of the manuscript appeared in the proceedings of the 21st International Conference on Web and Internet Economics (WINE’25).}
}

\author{
Sumit Goel\thanks{Department of Economics, Ashoka University, India; 
\href{mailto:sumitgoel58@gmail.com}{sumitgoel58@gmail.com}; \href{https://orcid.org/0000-0003-3266-9035}{0000-0003-3266-9035}}
\quad
Yuki Tamura\thanks{CREST, École Polytechnique, IP Paris, France;
\href{mailto:yuki.tamura@polytechnique.edu}{yuki.tamura@polytechnique.edu}}
}

\maketitle        

\begin{abstract}
The Top Trading Cycles (TTC) rule is fundamental to object reallocation: it is the only rule that is individually rational, pair efficient, and strategyproof on the unrestricted domain. Despite its desirable properties, at some profiles, TTC precludes longer trading cycles that could lead to more appealing allocations. We explore restricted preference domains as an avenue for finding alternative rules satisfying the three axioms. Our main contribution is to introduce a richness condition on domains, the \df{top-two condition}, and show that it characterizes \df{TTC domains}: domains to which the above characterization of TTC extends. The condition requires that, within every subset of objects, any two objects that can each be ranked first can also be ranked as the top two, in both possible orders. For every domain that fails the top-two condition, we construct a rule distinct from TTC that is individually rational, Pareto efficient, and strategyproof, and that executes longer trading cycles that TTC precludes.
\end{abstract}


\section{Introduction}

This paper studies the object reallocation problem, introduced by \citet*{shapley1974cores}. Each agent is endowed with an indivisible object and has a strict preference over all objects. A rule maps reported preferences into a reallocation of the objects.\\

A fundamental rule for this problem is Top Trading Cycles (TTC). \citet*{shapley1974cores} proposed TTC, credited to David Gale, as an algorithm for finding an allocation in the core, and \citet*{roth1977weak} later proved that it produces the unique core allocation. Focusing on incentives, \citet*{roth1982incentive} showed that TTC is strategyproof. In a seminal result, \citet*{ma1994strategy} characterized TTC as the unique rule satisfying individual rationality, Pareto efficiency, and strategyproofness. Alternative proofs were later provided by \citet*{svensson1999strategy, anno2015short, sethuraman2016alternative, bade2019matching}. More recently, \citet*{ekici2024pair} strengthened this characterization by replacing Pareto efficiency with pair efficiency, a substantially weaker requirement that only rules out mutually beneficial swaps between two agents. \citet*{ekici2024characterizing} provide a short proof of this result. Alternative characterizations using axioms such as group strategyproofness \citep*{bird1984group, takamiya2001coalition}, independence of irrelevant rankings \citep*{morrill2013alternative}, non-bossiness \citep*{miyagawa2002strategy, ehlers2014top}, and endowments-swapping-proofness \citep*{fujinaka2018endowments} have been established.\footnote{Some related strands of literature focus on reallocation problems tailored to specific environments \citep*{abdulkadirouglu1999house, roth2004kidney, schummer2013assignment} and object allocation problems \citep*{carroll2014general, hylland1979efficient, papai2000strategyproof, pycia2017incentive}. \citet*{morrill2024top} survey this literature and highlight the relevance of TTC in these environments.}\\

Despite these strong properties, TTC allocations may sometimes be less appealing
than alternative allocations. To illustrate, consider the profile in
Table~\ref{tab:ttc-precludes-long-cycle}. Under TTC, only agents 1 and 2 are
assigned their first-ranked objects, while all other agents receive their
second-ranked objects. However, there is an alternative allocation in which agent
1 instead receives their second-ranked object, and every other agent receives
their first-ranked object. The alternative moves $n-2$ agents from their
second- to their first-ranked object, at the cost of moving a single agent in the
opposite direction. This illustrates how TTC may preclude the execution of
arbitrarily long trading cycles, and raises the question of whether such
alternative allocations, which are arguably more desirable, can be selected while
retaining strategyproofness. A possible avenue is restricted preference domains,
which weaken the force of strategyproofness and potentially enrich the class of
rules satisfying the axioms.\\

\begin{table}[h]
\centering
\begin{tabular}{cccccccc}
$P_1$ & $P_2$ & $P_3$ & $\cdots$ & $P_k$ & $\cdots$ & $P_{n-1}$ & $P_n$  \\\hline
$o_2$ & $o_1$ & $o_2$ & $\cdots$ & $o_{k-1}$ & $\cdots$ & $o_{n-2}$ & $o_{n-1}$  \\
$o_n$ & $o_2$ & $o_3$ & $\cdots$ & $o_k$ & $\cdots$ & $o_{n-1}$ & $o_n$  \\
$\vdots$ & $\vdots$ & $\vdots$ & $\ddots$ & $\vdots$ & $\ddots$ & $\vdots$ & $\vdots$
\end{tabular}
\caption{A preference profile in which TTC precludes the execution of a longer trading cycle.}
\label{tab:ttc-precludes-long-cycle}
\end{table}

This motivates our central question: which domain restrictions preserve the
characterizations of TTC by \citet*{ma1994strategy} and \citet*{ekici2024pair}, and which admit other rules satisfying the same axioms? The existing literature has
examined a few specific cases, with mixed results:
\citet*{bade2019matching} shows that the characterizations do not extend to the
single-peaked domain, while \citet*{tamura2023object} and
\citet*{hu2024characterization} show that they do extend to the single-dipped domain. In
this paper, we take a general approach that not only provides a unifying
perspective on these findings, but also answers the question for every domain restriction. We introduce a richness condition on
domains, which we call the \df{top-two condition}. The condition requires that,
within every subset of objects, if two objects can each be ranked first, then they
can also be ranked as the top two objects in both possible orders. Our results establish that the characterizations extend to a domain if and only if it satisfies the top-two condition.\\

We first show that the top-two condition is sufficient: if a domain satisfies the top-two condition, then TTC is the only rule that is individually rational, pair efficient, and strategyproof. To illustrate the idea, consider again the profile in Table~\ref{tab:ttc-precludes-long-cycle}. If the domain allows agent $1$ to report a preference that keeps their first-ranked object unchanged while moving their endowment up to second rank, then individual rationality and pair efficiency imply that the rule must assign agent $1$ their first-ranked object. Strategyproofness then implies that the rule must do so at the original profile in Table~\ref{tab:ttc-precludes-long-cycle} as well, as is the case under TTC. More generally, at any profile, the condition ensures that agents in a top trading cycle can move their endowments just below their first-ranked objects. Such auxiliary profiles are a key ingredient in existing proofs of the characterizations of \citet*{ma1994strategy} and \citet*{ekici2024pair}. At these profiles, while Pareto efficiency directly implies that trading cycles must be executed, \citet*{ekici2024pair} and \citet*{ekici2024characterizing} exploit the richness of the unrestricted domain to establish the same conclusion under pair efficiency. In comparison, we show that this conclusion follows with no additional richness beyond the top-two condition, thereby identifying a broad class of domains on which both characterizations continue to hold.\\

Turning to necessity, for every domain that fails the top-two condition, we construct a rule distinct from TTC that is individually rational, Pareto efficient, and strategyproof. We first consider domains in which the failure occurs within the full set of objects, so that some agent, say agent $1$, cannot rank their endowment second below another object that can be ranked first, say $o_2$. Our rule deviates from TTC at profiles such as the one displayed in Table~\ref{tab:ttc-precludes-long-cycle}, where agents $1$ and $2$ rank each other's endowments first. Instead of letting these agents swap their endowments, the rule assigns agent $1$ their second-ranked object and executes a longer trading cycle, in which agent $2$ still receives $o_1$ and every other agent in the cycle receives their most-preferred object other than $o_1$. By carefully selecting the profiles at which this deviation occurs, we ensure that the resulting rule satisfies the axioms. Notably, at the profile in Table~\ref{tab:ttc-precludes-long-cycle}, this rule selects the alternative allocation described above. We then extend the construction to an arbitrary domain that fails the top-two condition, by applying this rule within a maximal subset of objects on which the condition fails, while assigning the remaining agents their TTC assignments.\\

Together, our results imply that the top-two condition characterizes \df{TTC domains}: domains to which the characterizations of \citet*{ma1994strategy} and \citet*{ekici2024pair} extend. The condition thus provides a unifying perspective on existing findings for specific domains and a simple criterion for classifying previously unexplored ones.\footnote{In a similar spirit, \citet*{alcalde1994top} propose the top-dominance criterion for two-sided matching, and \citet*{aswal2003dictatorial} introduce the notion of linked domains in social choice. In early work, \citet*{sen1969necessary} identify domain restrictions ensuring transitivity of the majority relation. A survey by \citet*{elkind2022preference} reviews a wide range of domain restrictions, some of which may be relevant in our context.} For instance, the single-dipped domain, previously studied by \citet*{tamura2023object, hu2024characterization}, and the partial-agreement domain, which contains all preferences consistent with a predefined partial order, satisfy the top-two condition, and are therefore TTC domains. In contrast, the single-peaked domain, previously studied by \citet*{bade2019matching, tamura2022crawler}, and the circular domain, previously studied in social choice by \citet*{kim1980special,sato2010circular}, fail the top-two condition, and are therefore not TTC domains.



\section{Model}
\label{sec:model}

\subsection*{Preliminaries}
Let $N=\{1,\dots, n\}$ be a finite set of agents. Let $O=\{o_1, \dots, o_n\}$ be a finite set of objects such that $o_i$ denotes agent $i$'s endowment. Agents have strict preferences over objects. We denote by $\P$ the set of all strict linear orders over $O$, and we let $\D \subset \P$ denote a preference domain. Let $P=(P_i)_{i\in N}\in \D^N$ denote a preference profile where $P_i\in \D$ is agent $i$'s preference. Following standard convention, for $S\subset N$, we let $P_S=(P_i)_{i\in S}$ and $P_{-S}=(P_i)_{i\in N\setminus S}$. For each $P_0 \in \D$, we denote by $R_0$ the ``at least as desirable as'' relation associated with $P_0$, i.e., for each pair $o,o' \in O$, $o\rR_0o'$ if and only if either $o\rP_0o'$ or $o = o'$. We refer to a set of agents, their endowments, and their preferences over these objects as an \df{economy}. \\

An \df{allocation} $x:N\to O$ is a bijection that assigns to each agent an object. Let $\X$ be the set of allocations. For each $x\in \X$ and each $i\in N$, we denote by $x_i\in O$ the assignment of agent $i$ at $x$. A \df{rule} $\f:\D^N \to \X$ associates with each profile $P \in \D^N$ an allocation. 

\subsection*{Axioms}
We now introduce some standard properties of allocations and rules.\\

Given $P\in \D^N$, an allocation $x\in \X$ is \df{individually rational} at $P$ if for each $i\in N$, $x_i\rR_io_i$. A rule $\f:\D^N \to \X$ is \df{individually rational} if for each $P\in \D^N$, $\f(P)$ is individually rational at $P$. \\

Given $P\in \D^N$, an allocation $x\in \X$ is \df{Pareto efficient} at $P$ if there is no allocation $y\in \X$ such that for each $i\in N$,  $y_i\rR_ix_i$ and for some $j\in N$, $y_j\rP_jx_j$. A rule $\f:\D^N \to \X$ is \df{Pareto efficient} if for each $P\in \D^N$, $\f(P)$ is Pareto efficient at $P$. Given $P\in \D^N$, an allocation $x\in \X$ is \df{pair efficient} at $P$ if there is no $i,j\in N$ such that  $x_j\rP_ix_i$ and $x_i\rP_jx_j$. A rule $\f:\D^N \to \X$ is \df{pair efficient} if for each $P\in \D^N$, $\f(P)$ is pair efficient at $P$. \\

A rule $\f:\D^N \to \X$ is \df{strategyproof} if for each $P\in \D^N$, there is no $i\in N$ and $P_i'\in \D$ such that $\f_i(P_i', P_{-i}) \rP_i \f_i(P)$. A rule $\f:\D^N \to \X$ is \df{group strategyproof} if for each $P\in \D^N$, there is no nonempty $S\subset N$ and $P'_S\in \D^S$ such that for each $i\in S$,  $\f_i(P'_S,P_{-S}) \rR_i \f_i(P)$ and for some $j\in S$, $\f_j(P_S', P_{-S}) \rP_j \f_j(P)$.

\subsection*{Top Trading Cycles}
We now describe the TTC rule. For any profile $P\in \P^N$, the TTC algorithm determines an allocation as follows:
\begin{enumerate}
    \item[(i)] Each agent points to the agent who owns their most-preferred object.
    \item[(ii)] In the resulting directed graph, there is at least one cycle. All agents belonging to cycles are assigned the objects they point to and leave the economy.
    \item[(iii)] If agents remain in the economy, repeat the procedure with the remaining agents and objects. Otherwise, terminate.
\end{enumerate}

We let $TTC(P) \in \X$ denote the allocation that results from running this algorithm at profile $P\in \P^N$. For any domain $\D\subset \P$, the \df{TTC rule} is the rule that assigns to each $P\in \D^N$ the allocation $TTC(P)$. It is well known that TTC satisfies all the properties defined above on the unrestricted domain. Hence, the same properties hold on every domain $\D\subset \P$.

\begin{fact}
\label{cor:TTC_is_good}
For any $\D\subset \P$, the TTC rule is individually rational, Pareto efficient, and group strategyproof.
\end{fact}

For the unrestricted domain $\D = \P$, there is, in fact, no other rule that satisfies even the weak combination of individual rationality, pair efficiency, and strategyproofness \citep*{ekici2024pair}. In general, we say that a domain
$\D \subset \P$ is a \df{TTC domain} if TTC is the only rule satisfying these three axioms on $\D$.\footnote{The terminology follows a common convention in the literature on domain restrictions, where a family of domains is identified by the class of rules it characterizes; see, for example, \citet*{aswal2003dictatorial} on dictatorial domains.} Our goal is to provide a general characterization of these domains.

\subsection*{Notation}

We will sometimes describe a preference $P_0\in \D$ by listing the objects in $O$ in the order specified by $P_0$. For example, $P_0\in \D$ such that $o_k\rP_0 o_{k+1}$ for all $k$ can be succinctly represented as $o_1o_2\dots o_n$.\\

For any $P_0\in \D$ and $O'\subset O$, we let $r_k\left(P_0, O'\right)$ denote the object ranked $k$-th under $P_0$ restricted to $O'$. For example, $r_1(P_0, O')$ is the object ranked first according to $P_0$ among $O'$.\\

Similarly, for any $\D'\subset \D$ and $O'\subset O$, we let $r_k(\D', O')$ denote the set of objects that are ranked $k$-th according to some $P_0\in \D'$ restricted to $O'$. Formally, 
$$r_k(\D', O')=\{o' \in O': \text{there exists } P_0\in \D' \text{ such that } r_k(P_0, O') = o'\}.$$
For example, $r_1(\D, O')$ is the set of objects ranked first according to preferences in $\D$ among $O'$.


\section{Results}
\label{sec:results}
\subsection{Top-two condition}
\label{subsec:top-two}

We now introduce our richness condition on preference domains. We first define the condition
within a fixed subset of objects. A domain satisfies the top-two condition within
$O'\subset O$ if any two objects that can be ranked first within $O'$ can also be ranked
as the top two within $O'$, in both possible orders.

\begin{definition}
\label{def:top-two-O}
A domain $\D\subset \P$ satisfies the \df{top-two condition within $O'\subset O$} if for
any distinct $a,b\in r_1(\D,O')$, there exists $P_0\in \D$ such that
$$
a=r_1(P_0,O')
\quad\text{and}\quad
b=r_2(P_0,O').
$$
\end{definition}

\begin{definition}
\label{def:top-two}
A domain $\D\subset \P$ satisfies the \df{top-two condition} if, for each $O'\subset O$,
$\D$ satisfies the top-two condition within $O'$.
\end{definition}

We now illustrate the condition with some examples. For any domain $\D\subset \P$, the
top-two condition within $O'$ is trivially satisfied whenever $|O'|\leq 2$, so it suffices
to check subsets of size at least three. Consider
$$
\D_1=\{o_1o_2o_3,\; o_2o_3o_1,\; o_2o_1o_3\}$$
and 
$$
\D_2=\{o_1o_2o_3,\; o_2o_3o_1,\; o_1o_3o_2\}.
$$
In both domains, only $o_1$ and $o_2$ can be ranked first within $O$. In $\D_1$, these
two objects can be ranked as the top two in both possible orders, so $\D_1$ satisfies the
top-two condition. In $\D_2$, however, no preference ranks $o_2$ first and $o_1$ second,
so $\D_2$ does not satisfy the top-two condition. The condition can also fail on a proper subset of objects. Consider
$$
\D_3=\{o_1o_2o_3o_4,\; o_1o_3o_2o_4,\; o_2o_1o_4o_3,\; o_2o_4o_3o_1\}.
$$
Although $\D_3$ satisfies the top-two condition within $O$, it fails the condition within
$O'=\{o_1,o_3,o_4\}$. Within this subset, both $o_1$ and $o_4$ can be ranked first, but no
preference ranks $o_4$ first and $o_1$ second. Thus, $\D_3$ fails the top-two condition.\\

The top-two condition is reminiscent of some existing richness conditions in the literature. One is \df{free pair at the top (FPT)}, which requires that any pair of objects $a,b\in O$ can be ranked as the top two \citep*{chatterji2011tops}. Another is Assumption B of \citet*{sonmez1999strategy}, which, adapted to our setting, requires that if a preference ranks $a$ above $b$, then it must be possible to move $b$ immediately below $a$, while leaving unchanged the sets of objects ranked above and below $a$.\footnote{\citet*{sonmez1999strategy} studies a more general object reallocation problem, in which agents may own multiple objects and have a preference over allocations, rather than only over their own assignments. Under a no-indifference condition and Assumption B on the preference domain, the paper establishes the significance of the core correspondence being (essentially) single-valued for existence of rules satisfying individual rationality, Pareto efficiency, and strategyproofness.} Both conditions are stronger than the top-two condition: any domain satisfying FPT or Assumption B satisfies the top-two condition, but the converse does not hold. For example, when $n\geq 3$, the singleton domain $\D=\{o_1o_2\dots o_n\}$ satisfies the top-two condition, but satisfies neither FPT nor Assumption B.

\subsection{Top-two is sufficient}
\label{subsec:sufficiency}

The following result shows that if a domain satisfies the top-two condition, it is a TTC domain.

\begin{thm}
\label{thm:SP_sufficient}
Suppose $\D\subset \P$ satisfies the top-two condition. Let $\f:\D^N\to \X$ be a rule. The following are equivalent:
\begin{enumerate}
    \item[(i)] $\f$ is the TTC rule.
    \item[(ii)] $\f$ is individually rational, Pareto efficient, and strategyproof.
    \item[(iii)] $\f$ is individually rational, pair efficient, and strategyproof.
\end{enumerate}
\end{thm}

To prove this result, we show that if $\f$ satisfies (iii), then at any profile $P\in \D^N$, it must coincide with TTC. Consider first the agents involved in a trading cycle in the first round of TTC. Since $\D$ satisfies the top-two condition, these agents can report their endowments as their second-ranked objects. At this auxiliary profile, individual rationality implies that $\f$ must either execute the cycle or assign all these agents their endowments.\\

While Pareto efficiency would immediately imply that $\f$ must execute the cycle, the core of our argument is to show that the same conclusion follows under the weaker requirement of pair efficiency (Lemma \ref{lem:helper}). Suppose the cycle is $S=\{i_1,\dots,i_{|S|}\}$, as in Table \ref{tab:ttc-key_profile}. Focus on agent $i_1$. We show that $\f$ must assign agent $i_1$ object $o_{i_k}$ if they instead report a preference $P^k_{i_1}\in\D$ that ranks $o_{i_k}$ first and $o_{i_{k+1}}$ second, as in Table \ref{tab:ttc-pair}. The base case $k=|S|$ follows directly from individual rationality and pair efficiency. For the induction step, assuming the claim holds for $k+1$, strategyproofness implies that when $i_1$ reports $P^k_{i_1}$, $\f$ must assign $i_1$ either $o_{i_k}$ or $o_{i_{k+1}}$. Pair efficiency rules out $o_{i_{k+1}}$, so $i_1$ must receive $o_{i_k}$. In particular, $\f$ assigns $i_1$ object $o_{i_2}$ when they report $P^2_{i_1}$. By strategyproofness, $\f$ must assign $o_{i_2}$ to $i_1$ at the auxiliary profile, and individual rationality then implies that $\f$ must execute the cycle $S$.\\

Finally, strategyproofness and individual rationality imply that $\f$ must execute the cycle at the original profile $P$ as well. Since the argument does not depend on the reports of agents outside the cycle, we can remove these agents and their assigned objects and apply the same argument to the agents in the next TTC cycle, as $\D$ satisfies the top-two condition within the reduced object set. Iterating over subsequent cycles yields $\f(P)=TTC(P)$. We now give the formal proof.

\begin{proof}[Proof of Theorem \ref{thm:SP_sufficient}]
The implications $(i)\implies (ii)\implies (iii)$ are immediate, so it suffices to prove $(iii)\implies (i)$. \\

Let $\f:\D^N \to \X$ be individually rational, pair efficient, and strategyproof. Fix $P\in \D^N$ and let $x=TTC(P)$. We show that $\f(P)=x$. \\

We first show that $\f$ executes the cycle in the first round of TTC at $P$.
Let $S\subset N$ be the set of agents in this cycle. If $|S|=1$, individual rationality immediately gives
$\f_i(P)=x_i$ for the unique $i\in S$. Hence assume $|S|\geq 2$. For any $i\in S$, both $x_i, o_i \in r_1(\D, O)$, and since $\D$ satisfies the top-two condition, there exists $P_i'\in \D$ such that $$x_i =r_1(P_i', O) \quad\text{and}\quad o_i =r_2(P_i', O).$$

Consider the profile $(P_S', P_{-S})$. By individual rationality, it must be that either  
\begin{enumerate}
    \item $\f_i(P'_S, P_{-S})=x_i$ for all $i\in S$, or 
    \item $\f_i(P'_S, P_{-S})=o_i$ for all $i\in S$.
\end{enumerate}

Were $\f$ Pareto efficient, we would immediately have that $\f$ executes the cycle. The following lemma shows that pair efficiency suffices under individual rationality and strategyproofness.

\begin{lem}
\label{lem:helper}
$\f_i(P'_S, P_{-S})=x_i$ for all $i\in S$.
\end{lem}
\begin{nestedproof}
If $|S|=2$, the claim follows immediately from pair efficiency, so assume $|S|\geq 3$. Label the agents $S=\{i_1,i_2,\ldots,i_{|S|}\}$ so that, for each
$k\in\{1,\ldots,|S|\}$, agent $i_k$ most prefers the endowment of agent $i_{k+1}$, where we adopt the cyclic convention
$i_{|S|+1}=i_1$.

\begin{table}[h]
\centering
\begin{tabular}{ccccccc}
$P'_{i_1}$ & $P'_{i_2}$ & $\cdots$ & $P'_{i_k}$ & $\cdots$ & $P'_{i_{|S|-1}}$ & $P'_{i_{|S|}}$  \\\hline
$o_{i_2}$ & $o_{i_3}$ & $\cdots$ & $o_{i_{k+1}}$ & $\cdots$ & $o_{i_{|S|}}$ & $o_{i_1}$  \\
$o_{i_1}$ & $o_{i_2}$ & $\cdots$ & $o_{i_k}$ & $\cdots$ & $o_{i_{|S|-1}}$ & $o_{i_{|S|}}$   \\
$\vdots$ & $\vdots$ & $\ddots$ & $\vdots$ & $\ddots$ & $\vdots$ & $\vdots$
\end{tabular}
\caption{Agents in cycle $S=\{i_1, \dots, i_{|S|}\}$ rank their endowment second.}
\label{tab:ttc-key_profile}
\end{table}

We construct a sequence of profiles differing only in $i_1$'s preference. For $k\in \{2, 3, \dots, |S|\}$, both $o_{i_k}, o_{i_{k+1}}\in r_1(\D, O)$, and since $\D$ satisfies the top-two condition, there exists $P^k_{i_1}\in \D$ such that $$o_{i_k} =r_1(P^k_{i_1}, O)\quad \text{and}\quad o_{i_{k+1}} =r_2(P^k_{i_1}, O).$$

\begin{table}[h]
\centering
\begin{tabular}{ccccccc}
$P^k_{i_1}$ & $P'_{i_2}$ & $\cdots$ & $P'_{i_k}$ & $\cdots$ & $P'_{i_{|S|-1}}$ & $P'_{i_{|S|}}$  \\\hline
$o_{i_k}$ & $o_{i_3}$ & $\cdots$ & $o_{i_{k+1}}$ & $\cdots$ & $o_{i_{|S|}}$ & $o_{i_1}$  \\
$o_{i_{k+1}}$ & $o_{i_2}$ & $\cdots$ & $o_{i_k}$ & $\cdots$ & $o_{i_{|S|-1}}$ & $o_{i_{|S|}}$   \\
$\vdots$ & $\vdots$ & $\ddots$ & $\vdots$ & $\ddots$ & $\vdots$ & $\vdots$
\end{tabular}
\caption{Agent $i_1$'s report used to move backward through the cycle using pair efficiency.}
\label{tab:ttc-pair}
\end{table}

Define the profile $P^k = (P_{i_1}^{k}, P'_{S\setminus \{i_1\}}, P_{-S})$.
We show that for every $k\in \{2, 3, \dots, |S|\}$, 
$$\f_{i_1}(P^k)=o_{i_k}.$$

First consider $k=|S|$. By individual rationality and pair efficiency, $\f_{i_1}(P^{|S|})=o_{i_{|S|}}$. Now consider any $k\in \{2, \dots, |S|-1\}$ and assume $\f_{i_1}(P^{k+1})=o_{i_{k+1}}$. Since $i_1$ can obtain $o_{i_{k+1}}$ by reporting $P^{k+1}_{i_1}$, strategyproofness implies that $\f_{i_1}(P^k) \in \{o_{i_k}, o_{i_{k+1}}\}$. But if $i_1$ gets $o_{i_{k+1}}$, by individual rationality, $i_k$ should get $o_{i_k}$, which violates pair efficiency. Therefore, $\f_{i_1}(P^k) = o_{i_k}$. \\

Thus, for every $k\in \{2, 3, \dots, |S|\}$, 
$\f_{i_1}(P^k)=o_{i_k}$. In particular, $\f_{i_1}(P^2) = o_{i_2}$. By strategyproofness, $\f_{i_1}(P'_S, P_{-S}) = o_{i_2}$. By individual rationality, $\f_i(P'_S, P_{-S})=x_i$ for all $i\in S$.
\end{nestedproof}

Returning to the proof of the theorem, Lemma \ref{lem:helper} gives $\f_i(P'_S, P_{-S})=x_i$ for all $i\in S$.  We now return from $P_S'$ to $P_S$ one agent at a time. Fix any $j\in S$. By strategyproofness, $\f_{j}(P_j, P'_{S\setminus \{j\}}, P_{-S}) = x_{j}$, and by individual rationality, for each $i\in S$, $$\f_{i}(P_j, P'_{S\setminus \{j\}}, P_{-S}) = x_{i}.$$

Now suppose for any $T\subset S$ where $|T|\leq k <|S|$, we have that  for each $i\in S$, $$\f_{i}(P_T, P'_{S\setminus T}, P_{-S}) = x_{i}.$$

Fix any $T$ of size $k+1$ and any $j\in T$. By strategyproofness, $\f_{j}(P_T, P'_{S\setminus T}, P_{-S}) = x_{j}$, and in fact, for any $i\in T$, $\f_{i}(P_T, P'_{S\setminus T}, P_{-S}) = x_{i}$. Now there must be some $i\in T$ such that $x_i=o_r$ where $r\in S\setminus T$. By individual rationality, for each $i\in S$, $\f_{i}(P_T, P'_{S\setminus T}, P_{-S}) = x_{i}$. It follows by induction that for each $i\in S$,
$$
\f_i(P_S,P_{-S})=\f_i(P)=x_i.
$$
Thus, $\f$ must execute the cycle in the first round of TTC at $P$.\\

We can now apply the same argument iteratively to show that $\f$ must also execute the subsequent cycles of the TTC algorithm. Note that the argument above does not depend on the reports of agents outside $S$, and hence, for every $Q_{-S}\in\D^{N\setminus S}$,
$$
\f_i(P_S,Q_{-S})=x_i
\quad\text{for all } i\in S.
$$
In words, $\f$ executes the cycle $S$ regardless of the reports of the remaining agents. This allows us to remove the agents in $S$ and their assigned objects, and apply the same argument to the reduced economy. The remaining objects form some subset $O'\subset O$.
Since $\D$ satisfies the top-two condition within $O'$, the argument applies with $O'$ in place of $O$. Iterating over all TTC cycles yields
$\f(P)=x=TTC(P)$. Since $P\in\D^N$ was arbitrary, $\f=TTC$.
\end{proof}

We now compare our proof of Theorem~\ref{thm:SP_sufficient} with existing proofs characterizing TTC on the unrestricted domain. To the best of our knowledge, all existing proofs, whether using Pareto efficiency \citep*{ma1994strategy}, or pair efficiency \citep*{ekici2024pair}, construct auxiliary profiles of the form displayed in Table~\ref{tab:ttc-key_profile}, where agents in a trading cycle rank their own endowment second. At such profiles, individual rationality and Pareto efficiency immediately imply that $\f$ must execute the cycle. A key challenge is to establish this using pair efficiency.\\

While the existing proofs with pair efficiency \citep*{ekici2024pair, ekici2024characterizing} differ in their underlying approaches, we focus on how they overcome this specific challenge and reinterpret their arguments in our context. \citet*{ekici2024pair} employs a contradiction argument. Assuming $\f$ satisfies the axioms but differs from TTC, there is some profile at which $\f$ does not execute a TTC cycle. Taking $S=\{i_1, \dots, i_{|S|}\}$ to be a smallest such cycle across profiles, the argument proceeds by showing that, at the auxiliary profile in Table~\ref{tab:ttc-key_profile}, $\f$ must assign each agent in $S$ their endowment. From here, if agent $i_1$ instead reports preferences whose top-three objects are $o_{i_2} o_{i_3} o_{i_1}$ and subsequently $o_{i_3} o_{i_1} o_{i_2}$, $\f$ must continue to assign $i_1$ their endowment. But then, in the second case, $\f$ fails to execute a smaller cycle of size $|S|-1$, contradicting the minimality of $S$. In a shorter proof, \citet*{ekici2024characterizing} show that $\f$ must execute the cycle $S$ at a profile of the form in Table~\ref{tab:ttc-key_profile}; otherwise, agent $i_1$ can instead report a preference $o_{i_2} o_{i_3} \cdots o_{i_{|S|}} o_{i_1}$, where individual rationality and pair efficiency imply that $\f$ must assign $o_{i_2}$ to $i_1$, thereby violating strategyproofness.\\

In comparison, our proof uses preferences of the form displayed in Table~\ref{tab:ttc-pair}, where agent $i_1$ reports $o_{i_k}$ and $o_{i_{k+1}}$ as their top two objects. These preferences require no additional richness of the domain beyond what is needed to construct the auxiliary profile in Table~\ref{tab:ttc-key_profile}. Thus, our proof uses minimal richness relative to the existing arguments. In fact, Theorem~\ref{thm:necessity} below shows that this richness cannot be weakened further. In addition, the direct approach of our proof illuminates the role played by each axiom in forcing the rule to coincide with TTC. The existing proofs have their own advantages: \citet*{ekici2024pair}'s proof offers a novel approach based on a notion of distance between mechanisms, and the proof of \citet*{ekici2024characterizing} does not exploit the details of the TTC rule, which may be valuable in other settings.\\

Theorem~\ref{thm:SP_sufficient} enables us to classify several important domains as TTC domains. We first consider the single-dipped domain, which has already been shown to be a TTC domain by \citet*{tamura2023object, hu2024characterization}.

\begin{example}[Single-dipped domain]
\label{ex:SD}
Suppose $\D^{SD}\subset \P$ contains preferences single-dipped with respect to the ordering $o_1 \to \dots \to o_n$:
    $$\D^{SD}=\{P_0\in \P: o_d = r_n(P_0, O) \implies o_{k}\rP_0o_{k+1} \text{ for } k<d \text{ and } o_{k+1}\rP_0o_{k} \text{ for } k \geq d\}.$$
For any $n\geq 2$ and any $O'\subset O$ with $|O'|\geq 2$, only the two extreme objects of $O'$ with respect to $o_1 \to \dots \to o_n$ can be ranked first within $O'$, and they can be ranked as the top two in both possible orders.  Thus, $\D^{SD}$ satisfies the top-two condition within every $O'\subset O$, and by Theorem \ref{thm:SP_sufficient}, is a TTC domain.
\end{example}

We next consider the partial-agreement domain, which, to the best of our knowledge, has not previously been explored in this context.\footnote{\citet*{nicolo2017age, fujinaka2024endowments} study a related domain, which they refer to as the common-ranking domain.} This domain captures environments in which there is an objective dominance relation over objects. For example, in kidney exchange, younger donor kidneys may dominate older donor kidneys, making preferences that violate this dominance relation implausible.

\begin{example}[Partial-agreement domain]
\label{ex:PA}
Suppose 
$\D^{PA}(\succ)\subset \P$ contains preferences consistent with the partial order $\succ$ over $O$:
    $$\D^{PA}(\succ) = \{P_0 \in \P: \text{ for all } a,b\in O,  a\succ b \implies a \rP_0 b \}.$$
     For any $n\geq 2$ and any $O' \subset O$ with $|O'|\geq 2$, only the undominated objects within $O'$ can be ranked first, and they can be ranked as the top two in both possible orders. Thus, for any partial order $\succ$ over $O$,  $\D^{PA}(\succ)$ satisfies the top-two condition within every $O'\subset O$, and by Theorem \ref{thm:SP_sufficient}, is a TTC domain. Note the two extreme cases: an empty partial order corresponds to the unrestricted domain $\D=\P$, while a complete order yields a domain containing a single preference, where individual rationality alone characterizes the TTC rule.
\end{example}

\subsection{Top-two is necessary}
\label{subsec:necessity}

We now turn to whether the top-two condition is also necessary for a domain to be a TTC domain. Equivalently, if a domain $\D\subset\P$ fails the top-two condition, must there exist a rule $\f:\D^N\to\X$, distinct from TTC, that satisfies the axioms in Theorem~\ref{thm:SP_sufficient}? The following result shows that this is indeed the case.

\begin{thm}
\label{thm:necessity}
Suppose $\D\subset\P$ does not satisfy the top-two
condition. Then there exists a rule
$\f:\D^N\to\X$, distinct from TTC, that is
individually rational, Pareto efficient, and
strategyproof.
\end{thm}

A natural approach is to construct such a rule directly. The difficulty is that the failure of the top-two condition provides only limited information about the domain: there exists a subset of objects within which two objects can each be ranked first, but cannot be ranked as the top two. Nevertheless, our proof of Theorem~\ref{thm:SP_sufficient} provides useful guidance for approaching the construction. Specifically, it establishes the following stronger claim: on \emph{any} domain $\D\subset\P$ and at \emph{any} profile $P\in\D^N$, any rule $\f$ satisfying the axioms must execute the TTC trading cycles one after another, for as long as any two objects in the current cycle can be ranked as the top two objects within the set of objects remaining at that stage. Thus, the first TTC trading cycle that $\f$ fails to execute must be one for which the failure of the top-two condition prevents the construction of the auxiliary preferences in Tables~\ref{tab:ttc-key_profile} and~\ref{tab:ttc-pair}.\\

We use this observation to construct such a rule in two steps. First, we consider the special case in which the top-two condition fails within the full object set $O$ (Lemma~\ref{lem:helper2}). In this case, there exist profiles at which two agents rank each other's endowments first, but one of them cannot report their endowment as their second-ranked object. On a carefully chosen nonempty subset of such profiles, our rule deviates from TTC by assigning this agent their second-ranked object instead of their TTC assignment, and executing a longer trading cycle in which the other agent still receives their first-ranked object. The choice of this subset ensures that the rule satisfies the axioms in the theorem.\\

In the second step, we use this construction to define a rule for an arbitrary domain that fails the top-two condition. We consider a maximal subset of objects within which the condition fails, where maximality ensures that every remaining agent can rank their endowment above all objects in this subset. When all of them do, our rule applies the first-step rule to the agents endowed with the objects in this subset. We now present the full proof.

\begin{proof}[Proof of Theorem \ref{thm:necessity}]
We first prove the result for the special case in which the top-two condition fails within the full object set.

\begin{lem}
\label{lem:helper2}
Suppose $\D\subset\P$ does not satisfy the top-two condition within $O$. Then there exists a rule $\f:\D^N\to\X$, distinct from TTC, that is individually rational, Pareto efficient, and strategyproof.
\end{lem}

\begin{nestedproof}
Since $\D$ does not satisfy the top-two condition within $O$, there exist two objects that can each be ranked first, but cannot be ranked as the top two in some order. Relabeling if necessary, assume that $o_1,o_2\in r_1(\D,O)$, and for all $P_0\in \D$, 
$$o_2 = r_1(P_0, O) \implies o_1 \neq r_2(P_0, O).$$

\noindent\textbf{Defining the rule $\f$}:

Our rule will deviate from TTC on a subset of profiles where agents $1$ and $2$ rank each other's endowments first, by assigning agent $1$ their second-ranked object, while still assigning $o_1$ to agent $2$. Since $o_1$ is then unavailable to every other agent, what matters for these agents is their most-preferred object other than $o_1$. Given any profile $P\in \D^N$, we say that a nonempty group $T\subset N\setminus\{1,2\}$ is \df{closed} at $P$ if $r_1(P_i,O\setminus\{o_1\})\in O_T$ for every $i\in T$.  Note that whether $T$ is closed at $P$ depends only on $P_T$. Define the following set of profiles:
$$
\Diff = \{ P \in \D^N :\; r_1(P_1, O) = o_2,\; r_1(P_2, O) = o_1, \text{ and no group } T\subset N\setminus\{1,2\} \text{ is closed at } P\}.
$$
Focusing on profiles where no group $T\subset N\setminus\{1,2\}$ is closed ensures that a longer trading cycle exists in which agent $1$ receives their second-ranked object instead of $o_2$. Indeed, starting from any agent in $N\setminus\{1,2\}$ and repeatedly moving to the owner of the current agent's most-preferred object other than $o_1$, one eventually reaches an agent whose most-preferred such object is $o_2$. Our rule deviates from TTC on profiles in $\Diff$ by executing this cycle.\\

Formally, we define our rule $\f : \D^N \to \X$ as follows:

$$\f(P) =
\begin{cases}
TTC(P) & \text{if } P \notin \Diff,\\
TTC\big(P_1{\downarrow}o_2,\; P_2,\; (P_i{\downarrow}o_1)_{i\in N\setminus\{1,2\}}\big) & \text{if } P \in \Diff,
\end{cases}
$$
where $P_0{\downarrow}o\in\P$ denotes the preference obtained from $P_0$ by moving $o$ to the bottom, leaving the relative ranking of all other objects unchanged. Note that $P_0{\downarrow}o$ need not belong to $\D$, but this causes no difficulty, since TTC is defined for every profile in $\P^N$.\\

To describe $\f$ on $\Diff$ more explicitly, fix any $P\in\Diff$. Let $o_{i_1}=r_2(P_1, O)$, and note that $i_1\notin\{1,2\}$, since no preference in $\D$ ranks $o_2$ first and $o_1$ second. Starting from $i_1$ and repeatedly moving to the owner of the current agent's most-preferred object other than $o_1$, we obtain a sequence of agents $i_1,\dots,i_m$ in $N\setminus\{1,2\}$ that ends with an agent $i_m$ such that $r_1(P_{i_m},O\setminus\{o_1\})=o_2$. No agent appears twice in this sequence, since the agents between two appearances would form a closed group. We refer to $C(P)=\{i_1,\dots,i_m\}$ as the \df{chain} at $P$, and note that it depends only on $P_1$ and on the preferences of the agents in $C(P)$. In the first round of TTC at the transformed profile, $1\to i_1\to i_2\to\dots\to i_m\to 2\to 1$ is a trading cycle. Letting $N'=N\setminus(\{1,2\}\cup C(P))$, it follows that
$$
\f_i(P)=
\begin{cases}
r_2(P_1,O) & \text{if } i=1,\\[2pt]
o_1 & \text{if } i=2,\\[2pt]
r_1(P_i,O\setminus\{o_1\}) & \text{if } i\in C(P),\\[2pt]
TTC_i(P_{N'}|O_{N'}) & \text{if } i\in N',
\end{cases}
$$
where the last expression uses that the transformed profile and $P$ coincide on $O_{N'}$.\\

\noindent\textbf{$\f$ satisfies the required properties:}

We first check that $\f$ is distinct from TTC. Since $o_1,o_2\in r_1(\D,O)$, there exists a profile $P\in \D^N$ with $r_1(P_2, O)=o_1$ and $r_1(P_i, O)=o_2$ for every $i\neq 2$. This profile belongs to $\Diff$, since every agent in $N\setminus\{1,2\}$ most prefers $o_2$ among $O\setminus\{o_1\}$, so that no group is closed. Hence, $\Diff$ is nonempty. Further, for every $P\in\Diff$, $TTC_1(P)=o_2$, while $\f_1(P)=r_2(P_1,O)\neq o_2$, and so $\f$ is distinct from TTC.\\
 
We now verify that $\f$ is individually rational. It suffices to consider $P\in\Diff$. Under the transformed profile, the set of objects an agent weakly prefers to their endowment can only become smaller. Since TTC is individually rational, each agent receives one of these objects, implying $\f(P)$ is individually rational at $P$.\\


We now show that $\f$ is Pareto efficient. Again, it suffices to consider $P\in \Diff$. Let $x=\f(P)$, and suppose $y\in\X$ is such that $y_i\rR_i x_i$ for every $i\in N$. We show that $y=x$. Since $x_2=o_1$ is agent $2$'s first-ranked object, $y_2=o_1$. Hence, no agent $i\in C(P)$ receives $o_1$ under $y$, and since $x_i$ is their most-preferred object in $O\setminus\{o_1\}$, $y_i=x_i$. In particular, some agent in $C(P)$ receives $o_2$, so agent $1$ cannot receive $o_2$, the only object they prefer to $x_1$, and thus $y_1=x_1$. It follows that $y$ assigns the objects in $O_{N'}$ to the agents in $N'$, and since $TTC(P_{N'}|O_{N'})$ is Pareto efficient, $y_i=x_i$ for every $i\in N'$ as well. Thus, $y=x$.\\

Finally, we show that $\f$ is strategyproof. Fix an agent $i\in N$ and fix $P_{-i}\in\D^{N\setminus\{i\}}$. There are two possibilities. First, $P_{-i}$ may be such that no report of agent $i$ can make the profile belong to $\Diff$: $(P_i',P_{-i})\notin\Diff$ for every $P_i'\in\D$. In this case, $\f$ coincides with TTC for every report of agent $i$, and hence agent $i$ cannot gain by misreporting. Second, $P_{-i}$ may be such that some report of agent $i$ makes the profile belong to $\Diff$: there exists $P_i'\in\D$ such that $(P_i',P_{-i})\in\Diff$. We analyze this case separately for agent $1$, agent $2$, and the agents in $N\setminus\{1,2\}$.
 
\begin{enumerate}
        \item $i=1$: By definition of $\Diff$, $P_{-1}$ is such that $r_1(P_2,O)=o_1$, and no group $T\subset N\setminus\{1,2\}$ is closed. We claim that, for every $P_1\in\D$, $$\f_1(P_1,P_{-1})=r_1(P_1,O\setminus\{o_2\}).$$
    If $(P_1,P_{-1})\in\Diff$, then $r_1(P_1,O)=o_2$, and by the explicit description of $\f$, agent $1$ receives their second-ranked object in $O$, which is exactly $r_1(P_1,O\setminus\{o_2\})$.

    If $(P_1,P_{-1})\notin\Diff$, then $r_1(P_1,O)\neq o_2$, and $\f$ coincides with TTC. Consider the first round of TTC at $(P_1,P_{-1})$. No cycle can consist only of agents in $N\setminus\{1,2\}$, since these agents would form a closed group. Hence, starting from any agent in $N\setminus\{1, 2\}$ and following the agents they point to, we must reach agent $1$ or agent $2$. Since $r_1(P_2, O)=o_1$, agent $2$ points to agent $1$. Thus, starting from any agent in $N\setminus\{1\}$, we reach agent $1$, so agent $1$ belongs to a cycle in the first round of TTC and receives their first-ranked object in $O$, which is also $r_1(P_1,O\setminus\{o_2\})$.

    It follows that agent $1$ cannot gain by misreporting.
    \item $i=2$: For all $P\in \D^N$, $\f_2(P)=TTC_2(P)$. Thus, agent $2$ cannot gain by misreporting.
 
    \item $i\in N\setminus\{1,2\}$: By definition of $\Diff$, $P_{-i}$ is such that $r_1(P_1,O)=o_2$ and $r_1(P_2,O)=o_1$. Moreover, since some report of agent $i$ makes the profile belong to $\Diff$, no group that excludes $i$ can be closed. Thus, whatever agent $i$ reports, every closed group contains $i$.  Let $P_i\in\D$ denote agent $i$'s true preference, and let $P=(P_i,P_{-i})$. We consider two cases.

\begin{enumerate}
    \item $(P_i,P_{-i})\notin\Diff$.
We claim that
    $$\f_i(P_i,P_{-i})=r_1(P_i,O\setminus\{o_1\}).$$
Since $P\notin\Diff$, let $T\subset N\setminus \{1, 2\}$ be a group of minimal size that is closed at $P$. From above, $i\in T$. By definition, each agent in $T$ most prefers, among $O\setminus\{o_1\}$, an object owned by an agent in $T$, and by the minimality of $T$, these agents form a single cycle. Since agents $1$ and $2$ swap their endowments in the first round of TTC, the agents in $T$ trade along this cycle once $o_1$ has been assigned, and agent $i$ receives $r_1(P_i,O\setminus\{o_1\})$. Finally, there is no report where agent $i$ receives $o_1$, since agent $2$ receives it. Thus, agent $i$ cannot gain by misreporting in this case.

   \item  $(P_i,P_{-i})\in\Diff$. There are two subcases: $i\in C(P)$ or $i\notin C(P)$. 
    
    If $i\in C(P)$, then agent $i$ receives $r_1(P_i,O\setminus\{o_1\})$. As before, agent $i$ cannot receive $o_1$, and so they cannot gain by misreporting.
    
    Now suppose $i\notin C(P)$. Let $N'=N\setminus (\{1, 2\}\cup C(P))$, so that $i\in N'$. We claim that for every report $P_i'\in\D$,
    $$\f_i(P_i',P_{-i})=TTC_i(P_i',P_{N'\setminus\{i\}}|O_{N'}).$$

    If $(P_i',P_{-i})\in\Diff$, then the chain at $(P_i',P_{-i})$ is again $C(P)$, since $i\notin C(P)$, and the claim follows from the explicit description of $\f$.

    If $(P_i',P_{-i})\notin\Diff$, then by the argument in Case (a) applied to $P_i'$, there is a closed group $T$ of minimal size at $(P_i',P_{-i})$ such that $i\in T$, the agents in $T$ form a single cycle, and $\f_i(P_i',P_{-i})=r_1(P_i',O\setminus\{o_1\})$. Observe that $T\cap C(P)=\emptyset$. Indeed, if $T$ contained an agent of the chain, it would also contain the next agent of the chain, and so on, up to the last agent of the chain, who most prefers $o_2$ among $O\setminus\{o_1\}$. But $o_2$ is not owned by any agent in $T$. Hence, $T\subset N'$. It follows that the agents in $T$ form the same cycle in the first round of TTC in the economy with agents $N'$ and objects $O_{N'}$, so that $TTC_i(P_i',P_{N'\setminus\{i\}}|O_{N'})=r_1(P_i',O\setminus\{o_1\})$ as well. This proves the claim. 
    
    Since TTC is strategyproof in the economy with agents $N'$ and objects $O_{N'}$, agent $i$ cannot gain by misreporting.
\end{enumerate}
\end{enumerate}

Thus, $\f$ is strategyproof.
\end{nestedproof}

We now use the construction in Lemma \ref{lem:helper2} to prove the result. Since $\D$ does not satisfy the top-two condition, let $S\subset N$ be such that $O_S$ is a maximal subset of objects within which $\D$ fails the top-two condition. If $O_S=O$, then the rule from Lemma \ref{lem:helper2} satisfies the required properties, and we are done. So assume $O_S\subsetneq O$.\\

Our rule will deviate from TTC on a subset of profiles where every agent $j\in N\setminus S$ ranks $o_j$ above every object in $O_S$, in which case it assigns these agents their TTC outcomes, while using the rule from Lemma~\ref{lem:helper2} to assign the objects in $O_S$ to the agents in $S$. To formalize this, observe that for any $j\in N\setminus S$, $o_j\in r_1(\D, O_S\cup\{o_j\})$. Otherwise, $\D$ fails the top-two condition within $O_S\cup\{o_j\}$, contradicting the maximality of $O_S$. For each $j\in N\setminus S$, define
$$
\D_j=\{P_0\in\D:o_j=r_1(P_0,O_S\cup\{o_j\})\},
$$
as the set of preferences in $\D$ that rank $o_j$ above every object in $O_S$. By the observation above, each $\D_j\neq \emptyset$.  We define $\f:\D^N\to\X$ as follows. For each $P\in\D^N$ and $j\in N\setminus S$,
$$
\f_j(P)=TTC_j(P),
$$
and for each $i\in S$,
$$
\f_i(P)=
\begin{cases}
TTC_i(P) & \text{if } P_j\notin\D_j \text{ for some } j\in N\setminus S,\\[4pt]
\f'_i(P_{S}|O_S) & \text{if } P_j\in\D_j \text{ for all } j\in N\setminus S,
\end{cases}
$$
where $\f'$ is the rule from Lemma~\ref{lem:helper2} for the economy with agents $S$, objects $O_S$, and the preferences over $O_S$ induced by $\D$, which fail the top-two condition within $O_S$.\\

We now verify that $\f$ satisfies the required properties. Consider any profile $P\in \D^N$ such that $P_j\in\D_j$ for all $j\in N\setminus S$. Since every agent in $N\setminus S$ ranks their endowment above every object in $O_S$, these agents trade only among themselves under TTC. Thus, the TTC rule assigns them the objects in $O\setminus O_S$, and assigns $TTC(P_S|O_S)$ to the agents in $S$. In contrast, our rule assigns $\f'(P_S|O_S)$ to the agents in $S$, and since $\f'$ differs from TTC, so does $\f$. It is straightforward to verify that $\f$ is individually rational and Pareto efficient. We now show that $\f$ is strategyproof. For any agent $j\in N\setminus S$, by definition, $\f_j(P)=TTC_j(P)$ for all $P\in \D^N$. Thus, these agents cannot gain by misreporting. For any agent $i\in S$, the branch of $\f$ is determined entirely by the reports of agents in $N\setminus S$, and hence cannot be affected by their own report. Conditional on the branch, their assignment is determined either by TTC or by $\f'$, both of which are strategyproof. Hence, these agents also cannot gain by misreporting. Thus, $\f$ is strategyproof.
\end{proof}

Theorem \ref{thm:necessity} enables us to show that several important domains are not TTC domains. Two such domains are the single-peaked domain (Example~\ref{ex:single-peaked}), for which \citet*{bade2019matching} already proposed the crawler rule, and the circular domain (Example~\ref{ex:circular}), which has been studied in social choice \citep*{kim1980special,sato2010circular} but remains unexplored in this context.

\begin{example}[Single-peaked domain]
\label{ex:single-peaked}
Suppose $\D^{SP}\subset \P$ contains preferences single-peaked with respect to the ordering $o_1 \to \dots \to o_n$:
$$\D^{SP}=\{P_0\in \P: o_p = r_1(P_0, O) \implies o_{k+1}\rP_0o_{k} \text{ for } k<p \text{ and } o_k\rP_0o_{k+1} \text{ for } k \geq p   \}.$$
For any $n\geq 3$, consider $O'=\{o_1, o_2, o_3\}$. While $o_1, o_3\in r_1(\D^{SP}, O')$, there is no $P_0\in \D^{SP}$ such that $o_1=r_1(P_0, O')$, $o_3=r_2(P_0, O')$, so $\D^{SP}$ does not satisfy the top-two condition within $O'$. By Theorem \ref{thm:necessity}, it is not a TTC domain.
\end{example}

\begin{example}[Circular domain]
\label{ex:circular}
Suppose $\D^C\subset \P$ contains preferences described by a first-ranked object, and a clockwise or counterclockwise traversal along the cyclic order $o_1\to o_2 \to \dots \to o_n \to o_1$: 
    $$\D^C = \{P_0\in \P: o_p=r_1(P_0, O) \implies P_0 \in \{o_p\dots o_n o_1 \dots o_{p-1}, o_p\dots o_1o_n \dots o_{p+1}\} \}.$$
For any $n\geq 4$, consider $O'=\{o_1, o_2, o_3, o_4\}$. While $o_1, o_3\in r_1(\D^{C}, O')$, there is no $P_0\in \D^{C}$ such that $o_1=r_1(P_0, O')$, $o_3=r_2(P_0, O')$, so $\D^{C}$ does not satisfy the top-two condition within $O'$. By Theorem \ref{thm:necessity}, it is not a TTC domain.
\end{example}

We also present the one-pair exclusion domain, which contains all rankings except those that rank an ordered pair as top-two, and is maximal among domains that fail the top-two condition. 

\begin{example}[One-pair exclusion domain]
\label{ex:maximal-gap}
Suppose $\D^{-o_io_j\dots}\subset \P$ contains all preferences over $O$ except those that rank $o_i$ first and $o_j$ second:
$$
\D^{-o_io_j\dots}=\P \setminus
\{P_0\in\P: r_1(P_0,O)=o_i,\ r_2(P_0,O)=o_j\}.
$$
For any $n\geq 3$, $\D^{-o_io_j\dots}$ fails the top-two condition within $O$: both $o_i$ and $o_j$ can be ranked first, but there is no $P_0\in \D^{-o_io_j\dots}$ that ranks $o_i$ first and $o_j$ second. By Theorem~\ref{thm:necessity}, it is not a TTC domain. Moreover, the failure of the top-two condition occurs only within the full set $O$: for every proper subset $O'\subsetneq O$, the domain admits arbitrary rankings of objects within $O'$, and hence satisfies the top-two condition within $O'$.
\end{example}

To see that one-pair exclusion domains are maximal, consider any domain $\D$ that fails the top-two condition. Then there exist a subset $O'\subset O$ and two objects $o_i,o_j\in r_1(\D, O')$ such that no preference in $\D$ ranks $o_i$ first and $o_j$ second within $O'$. It follows that no preference in $\D$ ranks $o_i$ first and $o_j$ second within $O$, since any such preference would also rank $o_i$ first and $o_j$ second within $O'$. Therefore, $\D\subset\D^{-o_io_j\dots}$. Since each one-pair exclusion domain contains $n!-(n-2)!$ preferences, any domain with more than $n!-(n-2)!$ preferences must satisfy the top-two condition, and hence is a TTC domain by Theorem~\ref{thm:SP_sufficient}.\\

Finally, we note that for all domains failing the top-two condition, our construction in the proof of Theorem~\ref{thm:necessity} demonstrates the possibility of executing the longer trading cycles that TTC precludes. For example, consider $\D^{-o_2o_1\dots}$ and any profile of the form displayed in Table~\ref{tab:ttc-precludes-long-cycle}. At this profile, TTC lets agents $1$ and $2$ swap their endowments, which precludes the longer trading cycle involving all agents. In contrast, the rule constructed in the proof of Lemma~\ref{lem:helper2} executes this cycle: it assigns agent $1$ their second-ranked object $o_n$, while every other agent receives their first-ranked object. Thus, excluding a single pair of top-ranked objects from the unrestricted domain already makes it possible to select such allocations, while retaining individual rationality, Pareto efficiency, and strategyproofness.\\

Together, Theorems~\ref{thm:SP_sufficient} and~\ref{thm:necessity} imply that the top-two condition characterizes TTC domains. 

\begin{corollary}
\label{cor:base}
Suppose $\D\subset \P$ is a domain. The following are equivalent:
\begin{enumerate}
\item[(i)] $\D$ satisfies the top-two condition.
\item[(ii)] There is a unique individually rational, Pareto efficient, and strategyproof rule on $\D$.
\item[(iii)] There is a unique individually rational, pair efficient, and strategyproof rule on $\D$.
\end{enumerate}
\end{corollary}

Thus, on any domain, \citet*{ekici2024pair}'s characterization holds if and only if \citet*{ma1994strategy}'s characterization holds.\\

We have focused on domain restrictions common to all agents, but our analysis can be used to derive analogous results for settings in which domain restrictions are heterogeneous across agents. We illustrate this in the following remark, where we identify a sufficient condition under which \citet*{ma1994strategy}'s characterization holds. We further present an example in which \citet*{ma1994strategy}'s characterization holds, but that of \citet*{ekici2024pair} does not, so that the above equivalence does not extend to this more general framework.

\begin{remark}[Heterogeneous domain restrictions]
\label{rem:heterogeneous}
Suppose that each agent $i\in N$ has a preference domain $\D_i\subset\P$. If, for every $i\in N$, $\D_i$ satisfies the top-two condition and $o_i\in r_1(\D_i,O)$, then TTC is the only rule $\f:\D_1\times\cdots\times\D_n\to\X$ that is individually rational, Pareto efficient, and strategyproof.\footnote{We thank the Advisory Editor for this suggestion.} Indeed, these assumptions ensure that the auxiliary profile in Table \ref{tab:ttc-key_profile} can always be constructed, so the same argument using Pareto efficiency in the proof of Theorem \ref{thm:SP_sufficient} forces $\f$ to coincide with TTC.\\

Under these conditions, however, Pareto efficiency cannot be replaced by pair efficiency, because preferences in Table \ref{tab:ttc-pair} may not be available. To see this, let $n=3$ and consider
$$\D_1=\{o_1o_2o_3,o_2o_1o_3\},\quad 
\D_2=\{o_2o_3o_1,o_3o_2o_1\}, \quad \D_3=\{o_3o_1o_2,o_1o_3o_2\}.$$
Notice that for every $i$, $\D_i$ satisfies the top-two condition and contains a preference that ranks $o_i$ first. From above, \citet*{ma1994strategy}'s characterization holds. Nevertheless, the rule that always selects the endowment allocation is individually rational, pair efficient, and strategyproof. Moreover, at the profile $P_1=o_2o_1o_3$, $P_2=o_3o_2o_1$, and $P_3=o_1o_3o_2$, this rule differs from TTC. Thus, the analogous characterization of \citet*{ekici2024pair} does not hold. If the assumptions on the domains were strengthened to ensure that the auxiliary profiles in both Tables~\ref{tab:ttc-key_profile} and~\ref{tab:ttc-pair} could be constructed, then the characterization using pair efficiency would follow as well.
\end{remark}

\section{Conclusion}
\label{sec:conclusion}

We introduce the top-two condition on preference domains and show that it characterizes TTC domains. On every domain satisfying the condition, TTC is the only rule that is individually rational, pair efficient, and strategyproof. On every domain that fails it, we construct a rule distinct from TTC that is individually rational, Pareto efficient, and strategyproof. Our construction also demonstrates how domain restrictions can permit longer trading cycles that TTC precludes, such as the one illustrated in Table~\ref{tab:ttc-precludes-long-cycle}. An interesting direction for future research is to understand more generally the structure of rules satisfying these axioms on domains that fail the top-two condition. Further work could also investigate analogous conditions for other characterizations of TTC.


\newpage
\bibliographystyle{ecta}

\bibliography{refs}

\end{document}